\documentclass[10pt]{article}
\usepackage{moriond,epsfig,amssymb,amsmath}

\def\be{\begin{equation}}
\def\ee{\end{equation}}
\def\bea{\begin{eqnarray}}
\def\eea{\end{eqnarray}}

\begin{document}
\rightline{CFTP/10-006}
\vspace*{4cm}
\title{NEUTRALINO DARK MATTER DETECTION BEYOND THE MSSM}

\author{NICOL\'AS BERNAL}

\address{Centro de F\'isica Te\'orica de Part\'iculas (CFTP)\\
Instituto Superior T\'ecnico, Avenida Rovisco Pais, 1049-001 Lisboa, Portugal}

\maketitle\abstracts{
The addition of non-renormalizable terms involving the Higgs fields
to the MSSM ameliorates the little hierarchy problem of the
MSSM. For neutralino dark matter, new regions for which the relic abundance
of the LSP is consistent with WMAP (as the bulk region and the stop coannihilation region)
are now permitted. In this framework, we analyze in detail first the direct dark matter detection
prospects in a XENON-like experiment; then we study the
capability of detecting gamma-rays produced
in the annihilation of neutralino LSPs in the FERMI satellite.
}

\section{Introduction}
The smallness of the quartic Higgs coupling in the framework of the
minimal supersymmetric standard model (MSSM) poses a problem. The tree
level bound on the Higgs mass is violated, and
large enough loop corrections to satisfy the lower bound on the Higgs
mass suggest that the stop sector has rather peculiar features: at
least one of the stop mass eigenstates should be rather heavy and/or
left-right-stop mixing should be substantial \cite{Djouadi:2005gj}.

The situation is different if the quartic Higgs couplings are affected
by New Physics. If the New Physics appears at an energy scale that is
somewhat higher than the electroweak breaking scale, then its effects
can be parametrized by non-renormalizable (NR) terms. The leading
NR terms that modify the quartic couplings are \cite{BMSSM}:
\begin{equation}\label{eq:wdst}
W_{\rm BMSSM}=\frac{\lambda_1}{M}(H_uH_d)^2+\frac{\lambda_2}{M}{\cal
  Z}(H_uH_d)^2,
\end{equation}
where ${\cal Z}=\theta^2m_{\rm susy}$ is a SUSY-breaking spurion.
The first term in equation~\ref{eq:wdst} is supersymmetric, while the
second breaks supersymmetry. In the scalar potential, the
following quartic terms are generated:
\begin{equation}\label{eq:dstsp}
2\epsilon_1 H_uH_d(H_u^\dagger H_u+H_d^\dagger H_d)
+\epsilon_2(H_uH_d)^2,
\end{equation}
where
\begin{equation}
\epsilon_1\equiv\frac{\mu^*\lambda_1}{M},\ \ \
\ \ \ \epsilon_2\equiv-\frac{m_{\rm susy}\lambda_2}{M}.
\end{equation}
Let us note that, in some regions of the parameter space, these operators Beyond the MSSM (BMSSM) may destabilize the scalar potential \cite{Blum:2009na}.

One of the attractive features of the MSSM is the fact that the lightest SUSY
particle (LSP), usually the lightest neutralino, is a natural candidate for being the dark matter (DM) particle.
The effects of the NR operators are potentially important in the determination of regions fulfilling the WMAP DM relic density constraint \cite{Dunkley:2008ie}.
On the one hand, these operators give rise to a new interaction Lagrangian which contributes to the vertex of two higgsinos and one or two Higgs bosons.
This effect is relevant when the neutralino LSP has a significant component of higgsinos.
On the other hand, the uplift of the lightest Higgs mass could reopen regions giving rise to a relic density in agreement wit the WMAP results, but which were ruled out by the bounds over the Higgs mass.
The effect of these operators on the relic density was studied in detail in references \cite{Berg:2009mq,Bernal:2009hd,Bernal:2009jc}.
In addition, the consequences for baryogenesis and electric dipole moments in the BMSSM has been studied in references \cite{baryogenesis,Bernal:2009hd}.
We shall evaluate the detection perspectives for two different detection modes,
namely direct detection in a XENON-like experiment and gamma-
ray detection from DM annihilations in the galactic center for the FERMI mission.

\section{The model}
The BMSSM framework, if relevant to the little hierarchy problem that
arises from the lower bound on the Higgs mass, assumes a New Physics
scale at a few TeV. Since the new degrees of freedom at this scale are
not specified, the effect of the new threshold on the running of parameters
from a much higher scale {\it cannot} be rigorously taken into account. It
therefore only makes sense to study the BMSSM effects in a framework
specified at low energy.
Within this framework, we calculate the DM relic density,
and the direct and indirect detection prospects
in the presence of
the new $\epsilon_i$ couplings. We used a modified version of the code
{\tt micrOMEGAs} \cite{micromegas}, where we implemented the BMSSM
couplings, in order to calculate
the relic density as well as the cross-sections and decay channels relevant
for DM detection.
The leading $\epsilon_i$-induced corrections to the
spectrum, were implemented using the code
{\tt SuSpect} \cite{Djouadi:2002ze}.

\begin{figure}[ht]
\begin{center}
\psfig{figure=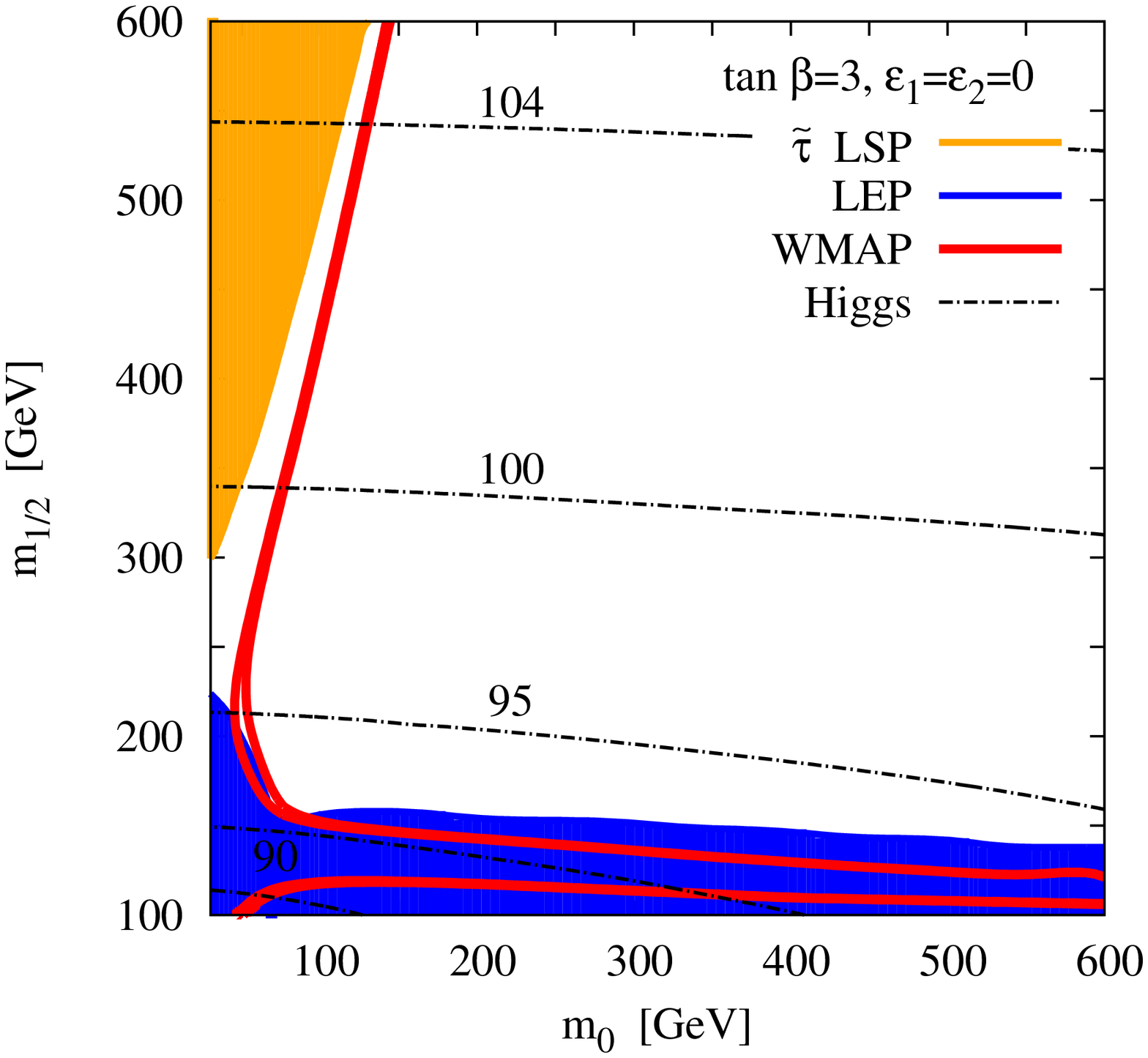  ,height=7cm,angle=-90}\hspace{-1cm}
\psfig{figure=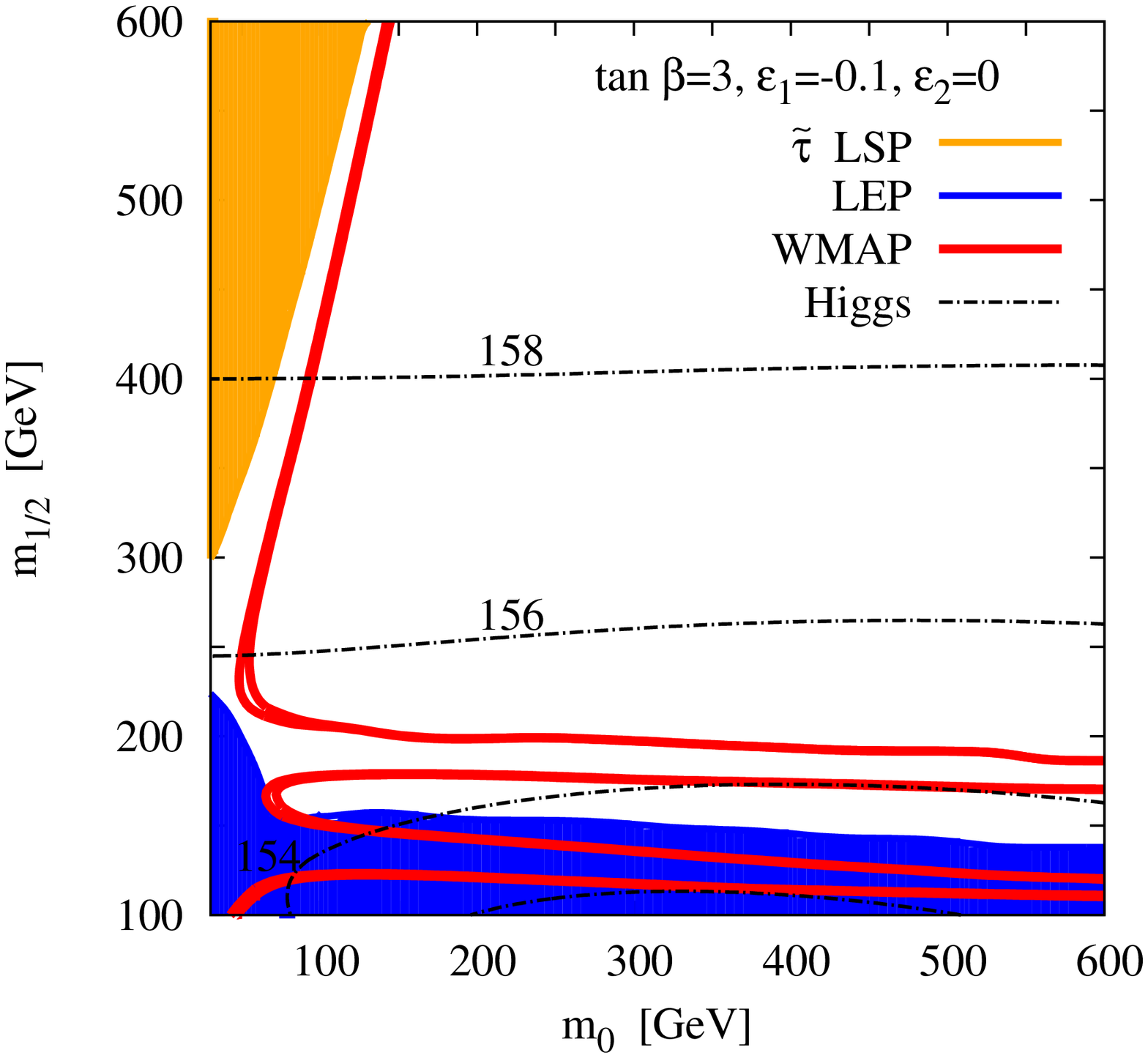,height=7cm,angle=-90}\vspace{-.3cm}
\psfig{figure=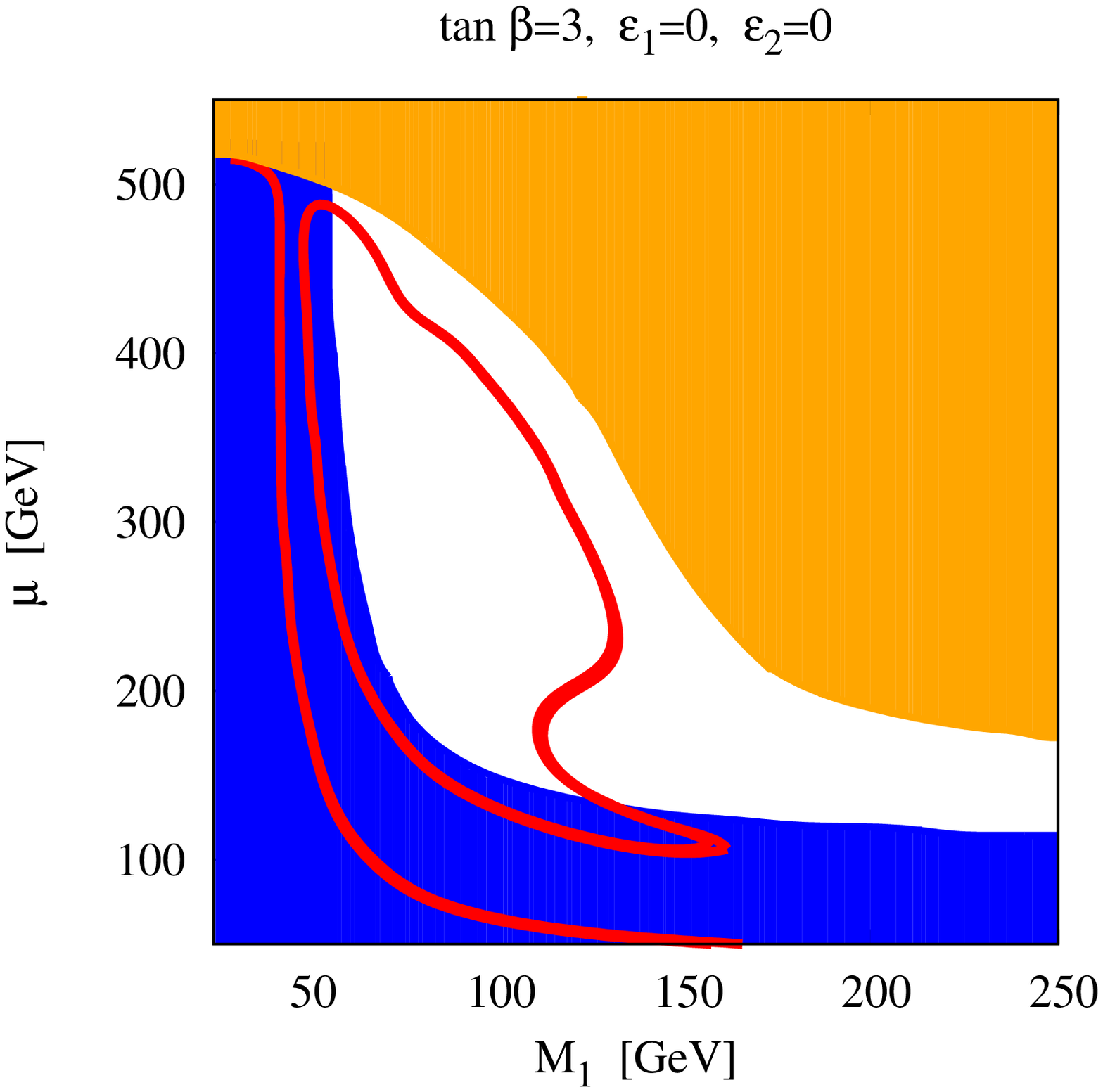  ,height=7cm,angle=-90}\hspace{-1cm}
\psfig{figure=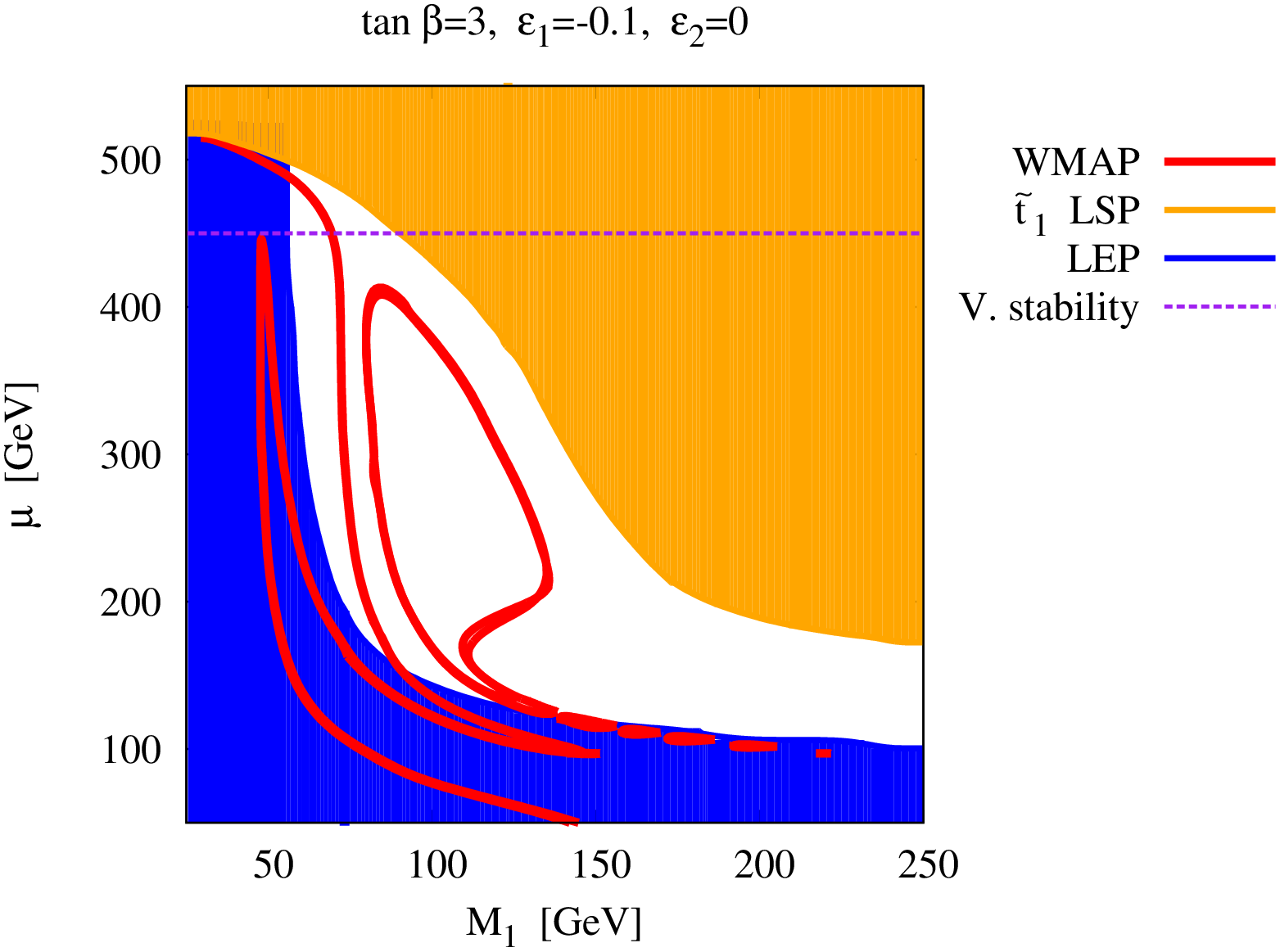,height=7cm,angle=-90}\vspace{-.3cm}
\caption{Regions in the $[m0,\,m_{1/2}]$ plane (upper panels) or $[M_1,\,\mu]$ plane (lower panels) in which the WMAP constraint is fulfilled (red lines).
The blue region is excluded by the null searches for charginos at LEP.
The orange areas in the upper (lower) panels are excluded because the stau (stop) is the LSP.
The dash-dotted black curves in the up panel are contour lines for the Higgs mass with values in GeV as indicated.
Above the dotted purple line of the left of this line, the electroweak vacuum is metastable.
\label{darkmatter}}
\end{center}
\end{figure}

The first scenario considered contains correlated stop and slepton masses,
as the mSUGRA framework. In this case,
the neutralino LSP is an almost pure bino-like state;
the `bulk region' is highly constrained
due to the experimental lower bound on the Higgs mass. In general, in order to fulfill such a
constraint either heavy or mixed stops are required.
Additionally to the ordinary mSUGRA parameters we have two
extra BMSSM parameters: $\epsilon_1$ and $\epsilon_2$.
Let us emphasize again that one should not think about this scenario as coming
from an extended mSUGRA model, since the effects of the BMSSM physics at the few TeV
scale on the running cannot be taken into account.
The upper panels of figure \ref{darkmatter} show the regions in the $[m_0,\,m_{1/2}]$ plane in which the WMAP constraint is fulfilled (red lines).
We fixed $A_0=0$ GeV, $\mu>0$, $\tan\beta=3$, $\epsilon_2=0$ and $\epsilon_1=0,\,-0.1$.
The blue region is excluded by the null searches for charginos at LEP.
The orange area is excluded because the stau is the LSP.
The dash-dotted black curves are contour lines for the Higgs mass.
Concerning the upper panels, a generic point in the former parameter space usually gives rise to a
  too large relic density, in conflict with the WMAP measurements.
However, for moderate $m_{1/2}$ and low $m_0$ values, there is a
  region where the LSP is almost degenerate in mass with the lightest stau,
  enhancing the co-annihilation cross-section
  $\chi_1^0-\tilde\tau$.
Another region giving rise to relic density in agreement with the
  WMAP measurements appears for $m_{1/2}\sim 120$ GeV (top left panel).
  This is the `$h$-pole' and the `$Z$-pole' region in which
  $m_h\sim m_Z\sim 2\,m_{\chi_1^0}$, and the $s$-channel Higgs and $Z$
  boson exchange is nearly resonant, allowing the
  neutralinos to annihilate efficiently.
  Let us note that this region is already excluded
  by LEP measurements.
For negative enough $\epsilon_1$ values (upper right panel), the uplift of the Higgs mass
  generates a splitting among the `$h$-pole' and the `$Z$-pole' regions, with the
  former now evading LEP constraints.

The second model correspond to a pure low energy scenario giving rise to light unmixed stops.
In addition to the BMSSM $\epsilon_i$
parameters, we consider the following set of parameters:
\begin{equation}
M_1,\ \mu,\ \tan\beta,\ X_t,\ m_U,\ m_Q, \ m_{\tilde f},\ m_A\,,
\end{equation}
where
$m_{\tilde f}$ is a common mass for the sleptons, the first and second
generation squarks, and $\tilde b_R$. We further use
$M_1=\frac53\,\tan^2\theta_W\,M_2$.
To demonstrate our main points, we fix the values of all but two
parameters as follows: $\epsilon_1=0$ or $-0.1$, $\epsilon_2=0$,
$\tan\beta=3$, $X_t=0$, $m_U=210$ GeV, $m_Q=400$ GeV,
$m_{\tilde f}=m_A=500$ GeV. This scenario gives rise to relatively
light stops, namely $m_{\tilde t_1}\lesssim 150$ GeV and $370$ GeV $\lesssim m_{\tilde t_2}\lesssim 400$ GeV.
We scan over the remaining two parameters, $M_1$ and $\mu$.

The lower panels of figure \ref{darkmatter} show the regions in the $[M_1,\,\mu]$ plane
in which the WMAP constraint is fulfilled (red lines). The first region is
the $Z$- and $h$-poles in which the LSP is rather light,
  and the $s$-channel $Z$ or $h$
  exchange is nearly resonant, allowing the neutralinos to annihilate efficiently.
There is also a `mixed region' in which the LSP is a higgsino--bino
mixture, $M_1 \sim\mu$, which enhances
(but not too much) its annihilation cross-sections into final states
containing gauge and/or Higgs bosons.
Finally the `stop co-annihilation' region, in which the LSP is
almost degenerate in mass with the lightest stop.
Such a scenario leads to an enhanced annihilation of sparticles
since the $\chi_1^0-\tilde t_1$ co-annihilation cross-section
is much larger than that of the LSP.

Let us first consider the case where $\epsilon_1=\epsilon_2=0$ (left lower panel).
In any case, for the Higgs mass values
obtained here, $m_h \sim 85$ GeV, this
region is already excluded by the negative searches for chargino pairs
at LEP.
The main difference when considering the $\epsilon_1=-0.1$ case (lower right panel) comes from the important enhancement of the Higgs mass
due to the presence of the BMSSM operators. In this case it is possible to
disentangle the $Z$ and the $h$ peaks, since the Higgs-related peak moves to
higher $M_1$ values, due to the increase of the Higgs mass: $m_h=122$
GeV. Let us note that the latter peak is no longer excluded
by chargino searches.

\section{Direct detection of dark matter}
DM direct detection experiments measure the number $N$ of elastic
collisions between WIMPs and target nuclei in a detector,
per unit detector mass and per unit of time, as a function of the
nuclear recoil energy $E_r$.
The differential event rate per unit detector mass and
per unit of time can be written as:
\begin{equation}
\frac{dN}{dE_r}=\frac{\sigma_0\,\rho_\odot}{2\,m_r^2\,m_\chi}\,
F(E_r)^2\int_{v_\text{min}(E_r)}^{\infty}\frac{f(v_\chi)}{v_\chi}dv_\chi\,,
\label{Recoil}
\end{equation}
where $\sigma_0$ is related to the
WIMP-nucleon cross-section, $\sigma_{\chi-p}$, by $\sigma_0=\sigma_{\chi-p}\cdot (A\,m_r/M_r)^2$, with $M_r=\frac{m_\chi\,m_p}{m_\chi+m_p}$ the WIMP-nucleon reduced mass, $m_r=\frac{m_\chi\,m_N}{m_\chi+m_N}$ the
WIMP-nucleus reduced mass, $m_\chi$ the WIMP mass, $m_N$ the nucleus mass,
and $A$ the atomic weight. $F$ is the nuclear form factor; in the following analysis the Woods-Saxon form factor will be used. $\rho_\odot\simeq0.385$ GeV cm$^{-3}$ and $f(v_\chi)$ are the density and the velocity distribution of WIMPs near the Earth.
Let us note that we are assuming identical WIMP-proton and WIMP-neutron cross-sections, and that we are ignoring the spin-dependent interactions.
In our study
we will consider a XENON detector with $7$ energy bins between $4$ and $30$ keV.
We also consider a negligible background.
Furthermore, we examine three `benchmark' experimental setups, assuming exposures
$\varepsilon=30$, $300$ and $3000$ kg$\cdot$year, which could correspond e.g.
to a detector with $1$ ton of xenon and $11$ days, $4$ months or $3$ years of data acquisition respectively.

\begin{figure}[ht]
\begin{center}
\psfig{figure=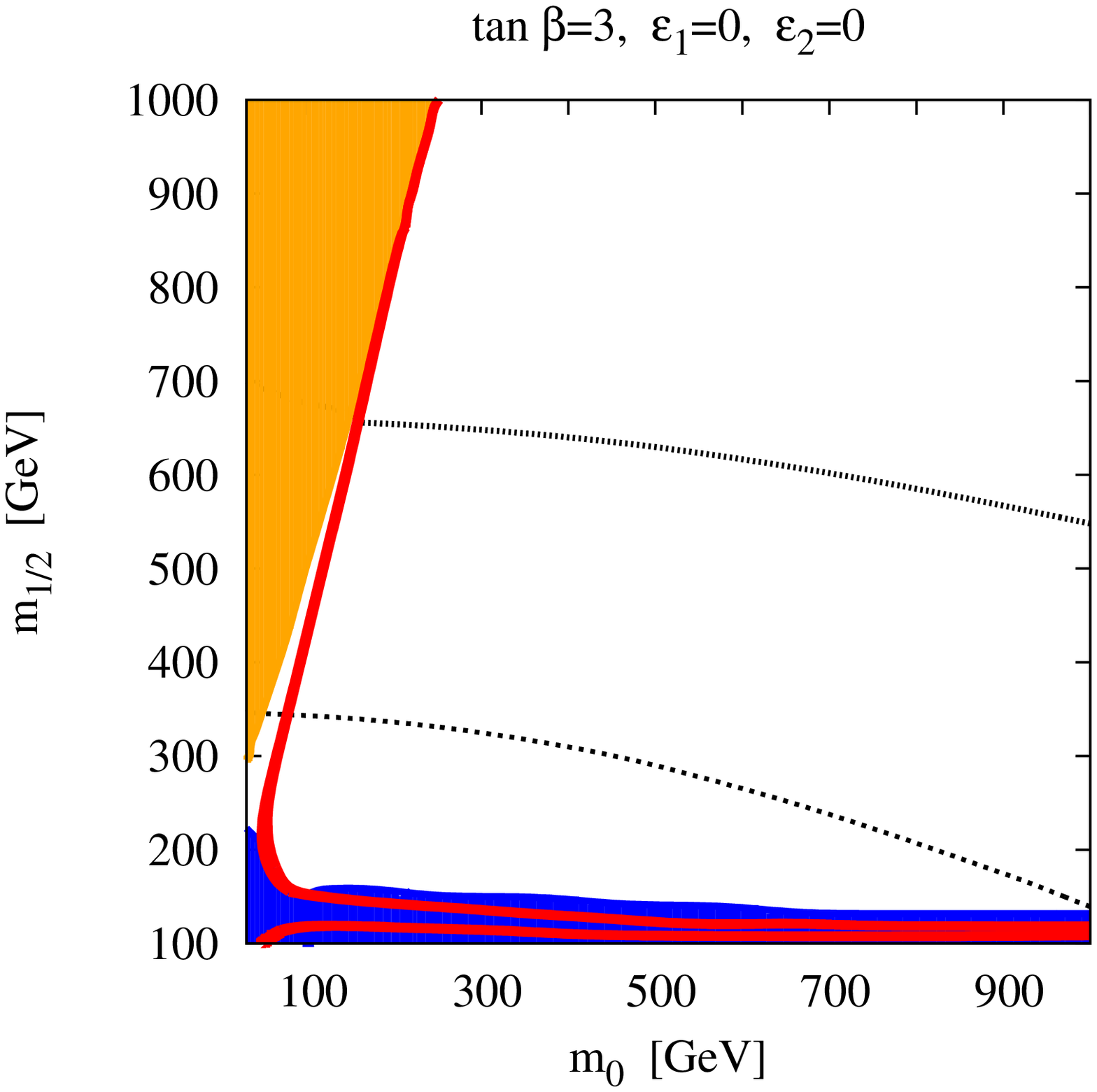  ,height=7cm,angle=-90}\hspace{-1cm}
\psfig{figure=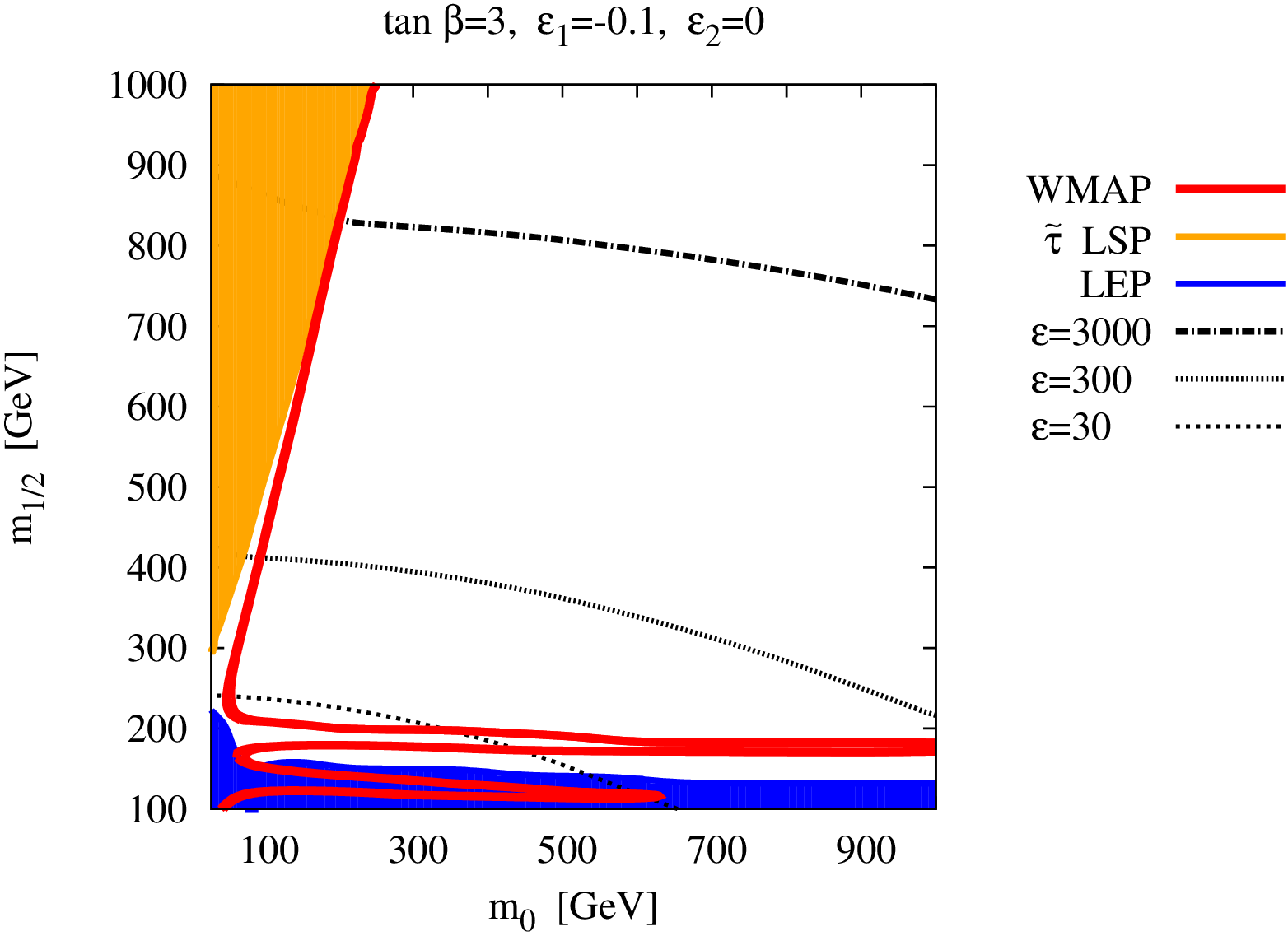,height=7cm,angle=-90}\vspace{-.3cm}
\psfig{figure=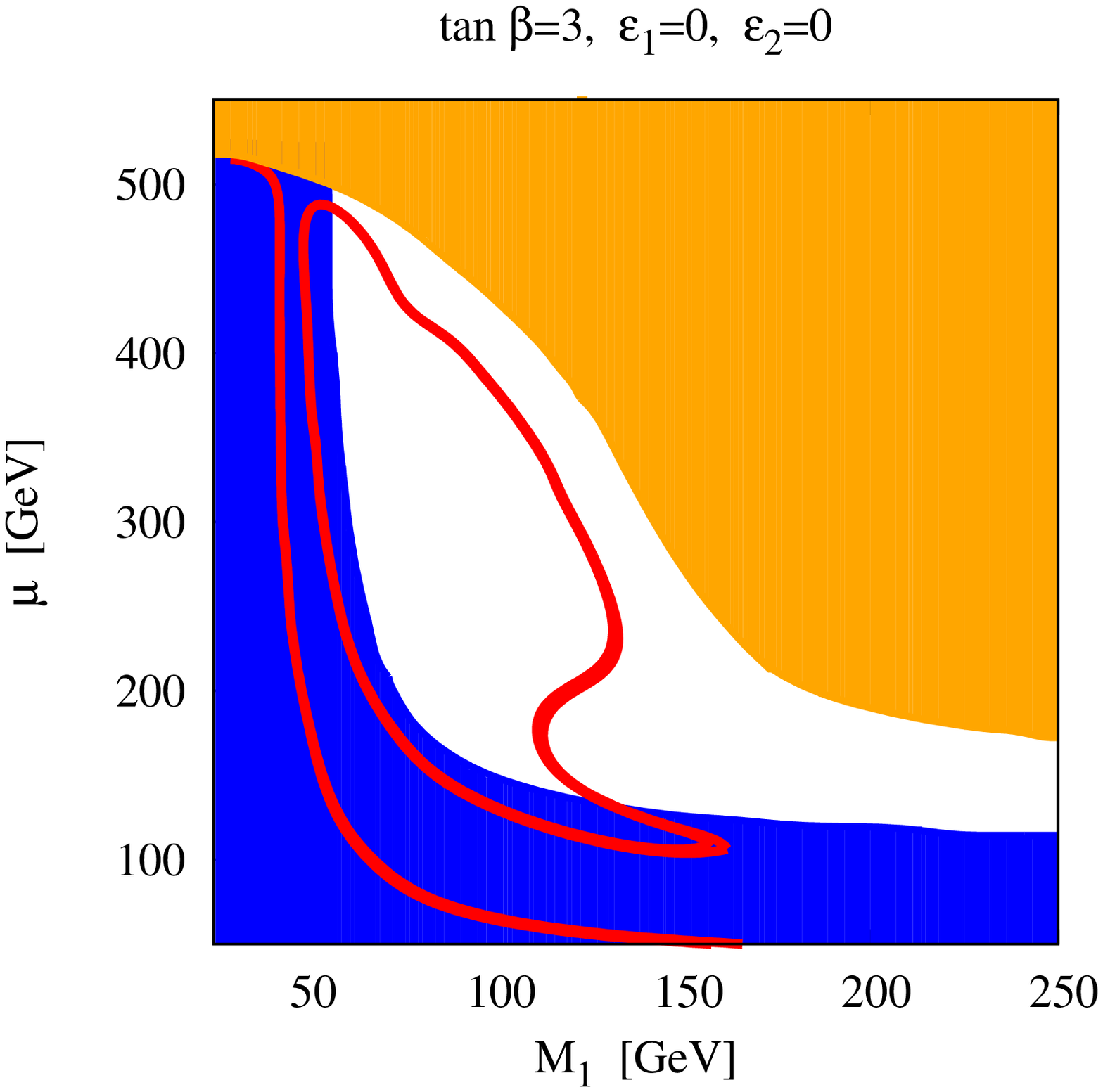  ,height=7cm,angle=-90}\hspace{-1cm}
\psfig{figure=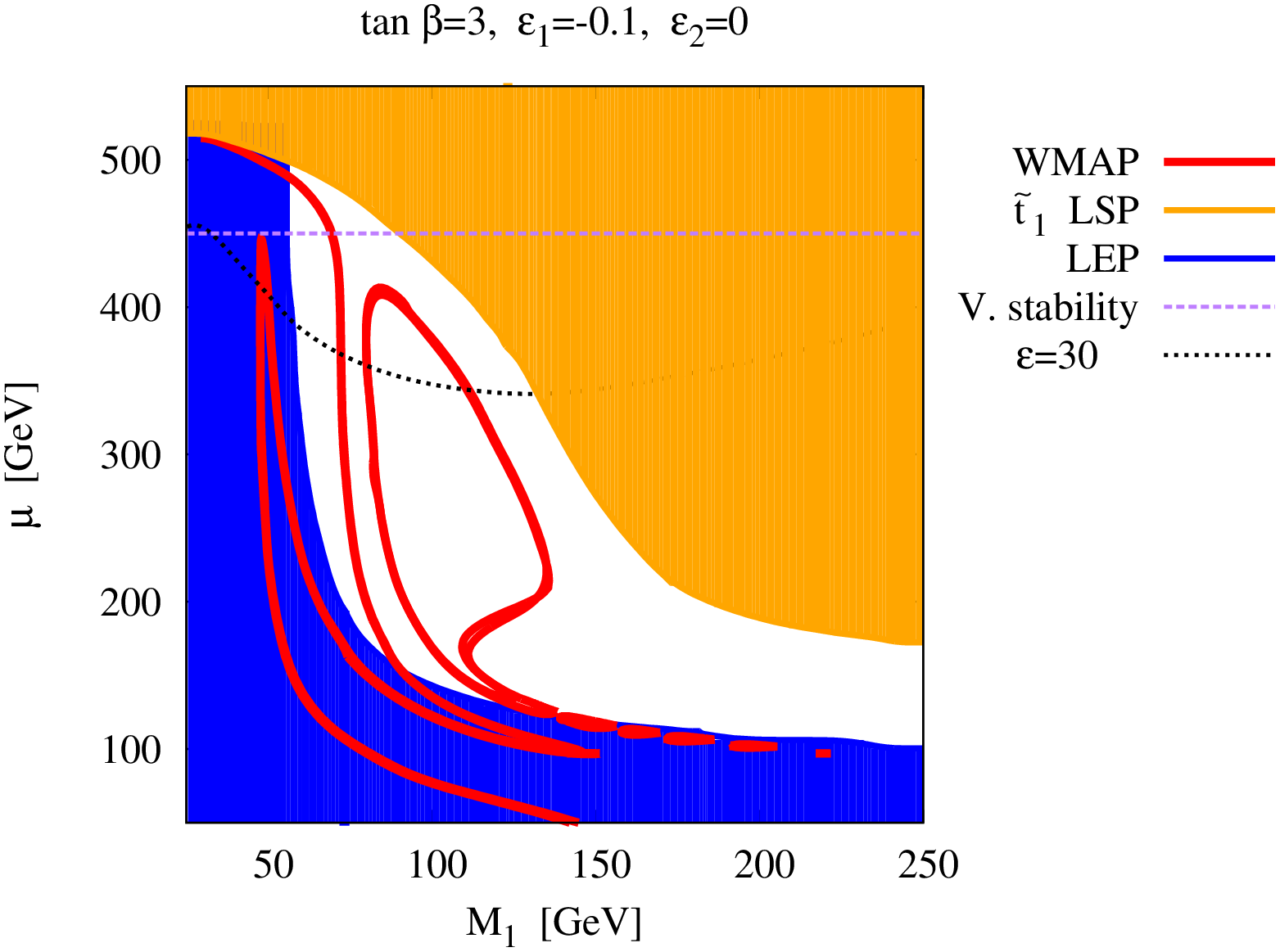,height=7cm,angle=-90}\vspace{-.3cm}
\caption{Regions in the $[m0,\,m_{1/2}]$ plane (upper panels) or $[M_1,\,\mu]$ plane (lower panels) in which the WMAP constraint is fulfilled (red lines).
The blue region is excluded by the null searches for charginos at LEP.
The orange areas in the upper (lower) panels are excluded because the stau (stop) is the LSP.
The dash-dotted black curves in the up panel are contour lines for the Higgs mass with values in GeV as indicated.
Above the dotted purple line of the left of this line, the electroweak vacuum is metastable.
\label{direct}}
\end{center}
\end{figure}

Figure \ref{direct} shows the exclusion lines (black lines) for
exposures $\varepsilon=30$, $300$ and $3000$ kg$\cdot$year,
on the $[m_0,\,m_{1/2}]$ (upper panels) or the $[M_1,\,\mu]$ (lower panels) parameter space for the two models defined previously.
The left plots correspond to plain MSSM scenarios whereas the right to the
BMSSM, with the $\epsilon_1$ parameter turned on.
These curves reflect the XENON sensitivity and represent its ability to test and exclude different regions of the parameter space at $95\%$ CL: all points lying below the lines are detectable.
We note that when some line is absent, this means that
the whole parameter space can be probed for the corresponding exposure

As a general rule, the detection prospects are maximised for low values of the pairs
$m_0$ and $m_{1/2}$ or $M_1$ and $\mu$ because they give rise to a light LSP.
On the other hand, the regions of low $m_{1/2}$ or $M_1\sim\mu$ are also preferred because in that case the
lightest neutralino is a mixed bino-higgsino state, favouring the $\chi_1^0-\chi_1^0-h$
and $\chi_1^0-\chi_1^0-H$ couplings, and therefore the scattering cross-section.
The detection prospects are also maximised for low values of $\tan\beta$.
The introduction of the NR operators gives rise to an important deterioration of the detection prospects.
The main effect enters via the important increase in the lightest CP-even Higgs mass.
Nevertheless, let us emphasize again that this deterioration is relative, since we are comparing
with a plain MSSM, which is already excluded because of the light Higgs mass.
Concerning the plots in figure \ref{direct}, a further remark that can be made is that,
even for low exposures, a sizable amount of the parameter space can be probed.
Larger exposures could be able to explore almost the whole parameter space taken
into account.

\section{Indirect detection of dark matter with gamma-rays}
The differential flux of gamma--rays generated from DM annihilations
and coming from a direction forming an angle $\psi$ with respect to
the galactic center (GC) is
\begin{equation}\label{Eq:flux}
\frac{d\Phi_{\gamma}}{dE_\gamma}(E_{\gamma}, \psi)
=\frac{\langle\sigma v\rangle}{8\pi\,m_{\chi}^2}\sum_i
\frac{dN_{\gamma}^i}{dE_{\gamma}}\,Br_i\,\int_\text{l.o.s.}\rho(r)^2\,dl\,,
\end{equation}
where the discrete sum is over all DM annihilation
channels,
$dN_{\gamma}^i/dE_{\gamma}$ is the differential gamma--ray yield of SM particles into photons,
$\langle\sigma v\rangle$ is the total self--annihilation cross-section averaged
over its velocity distribution, $\rho$ is the DM density profile, $r$
is the distance from the GC
and $Br_i$ is the branching ratio of annihilation into the $i$-th final state.
The integration
is performed along the line of sight from the observation point towards the GC.
In the following study we take into account three DM density profiles: NFW with and without adiabatic compression and the Einasto profile.
In our study
we will consider the FERMI satellite, taking into account a five-year mission run, and an energy range extending from $1$ up to $300$ GeV, with $20$ logarithmically evenly spaced bins.
We focus ourselves to an solid angle $\Delta\Omega=3\cdot 10^{-5}$ sr around the GC.
We will consider two sources for the high-energy gamma-ray background corresponding to the galactic emissions coming from resolved and diffuse sources. Both measurements has been done by the HESS collaboration.

\begin{figure}[ht]
\begin{center}
\psfig{figure=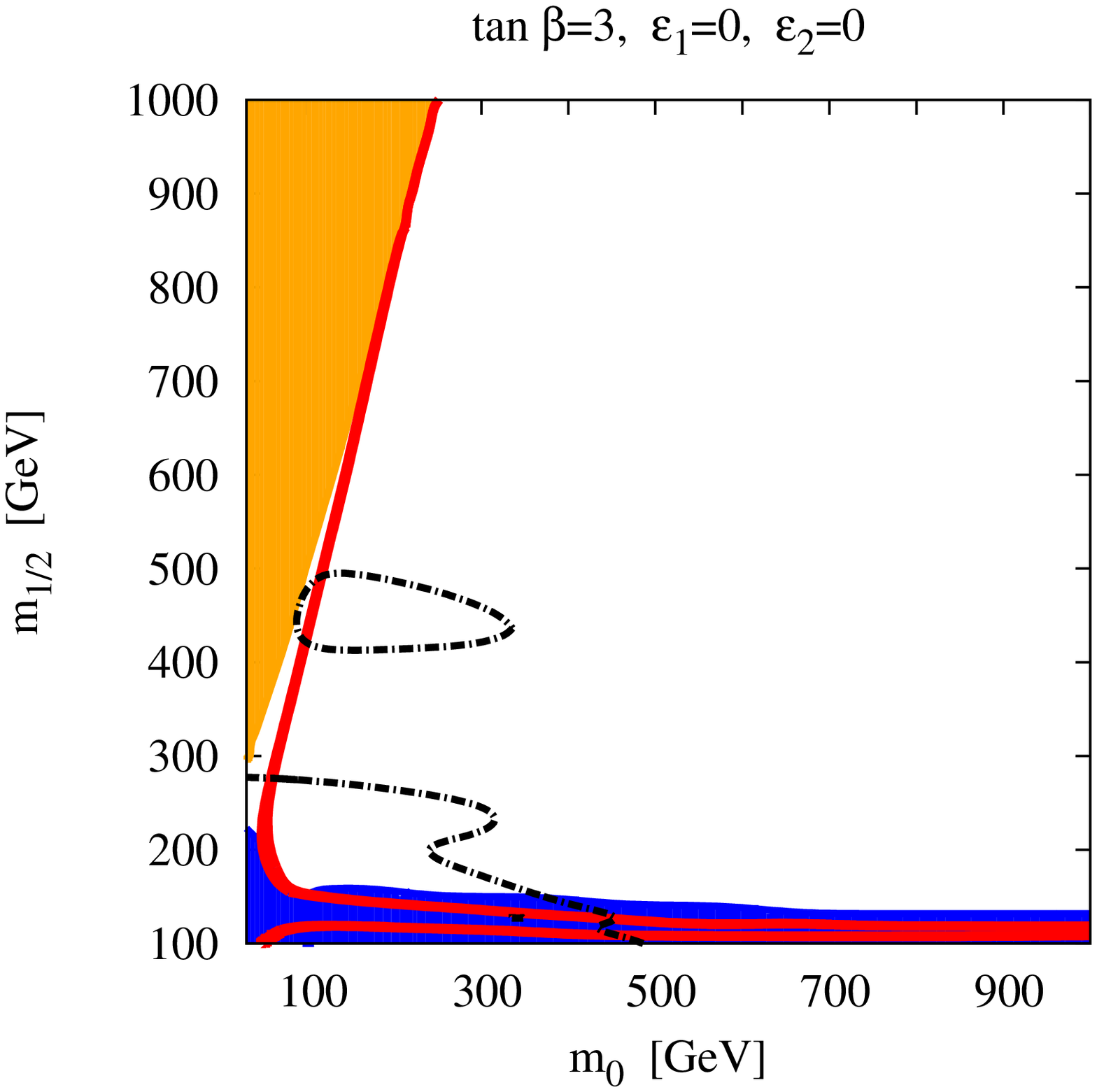  ,height=7cm,angle=-90}\hspace{-1cm}
\psfig{figure=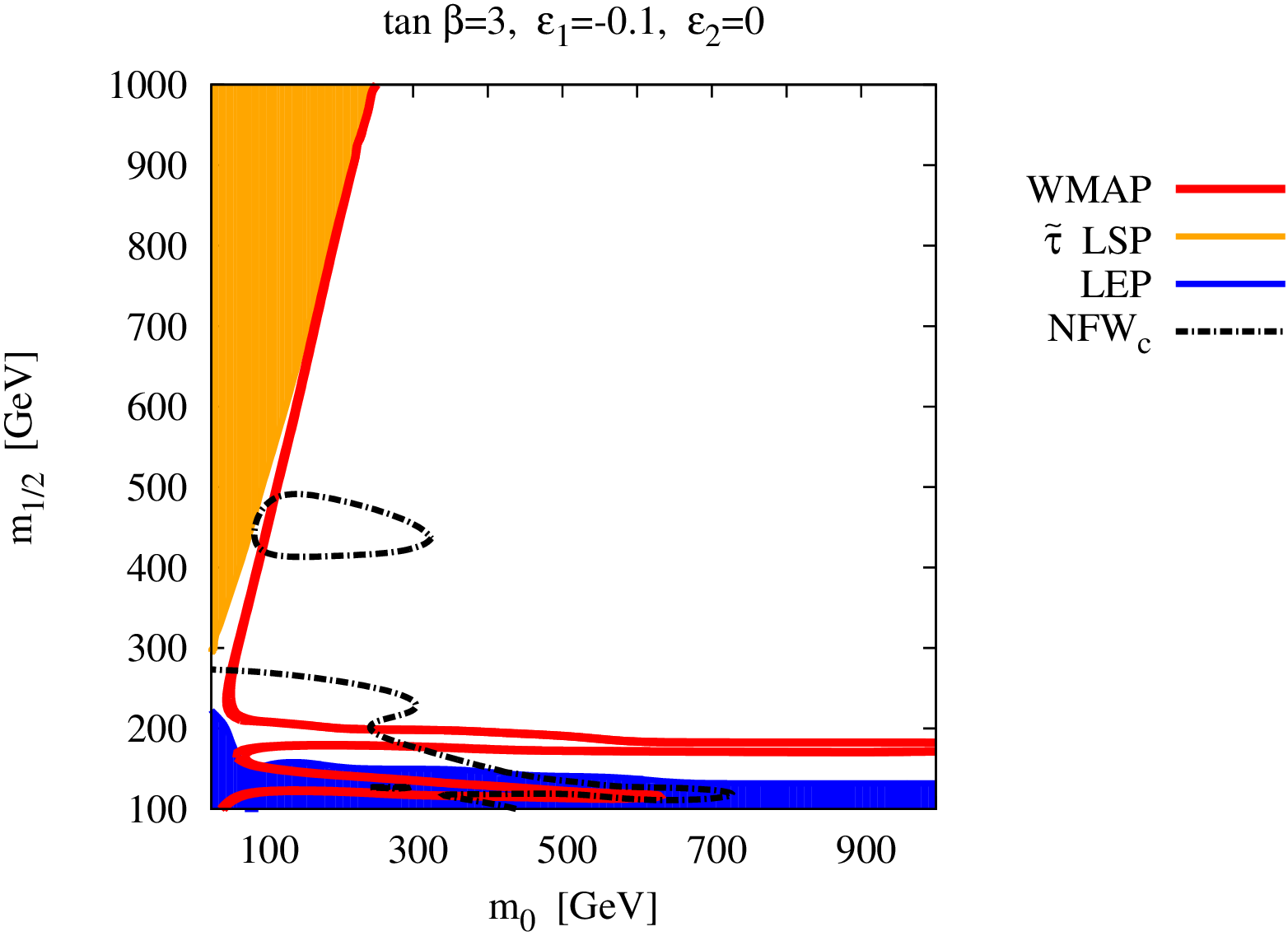,height=7cm,angle=-90}\vspace{-.3cm}
\psfig{figure=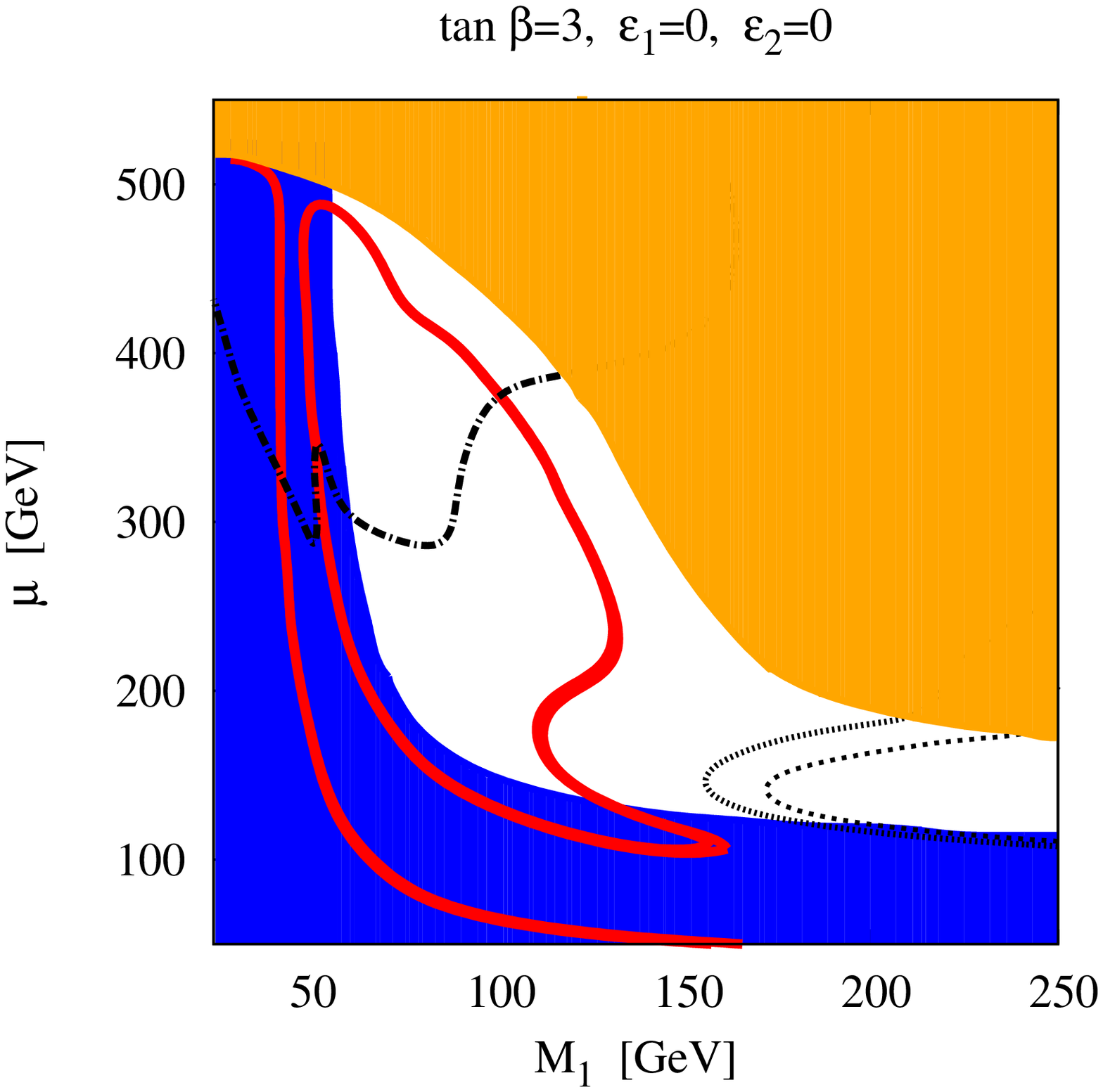  ,height=7cm,angle=-90}\hspace{-1cm}
\psfig{figure=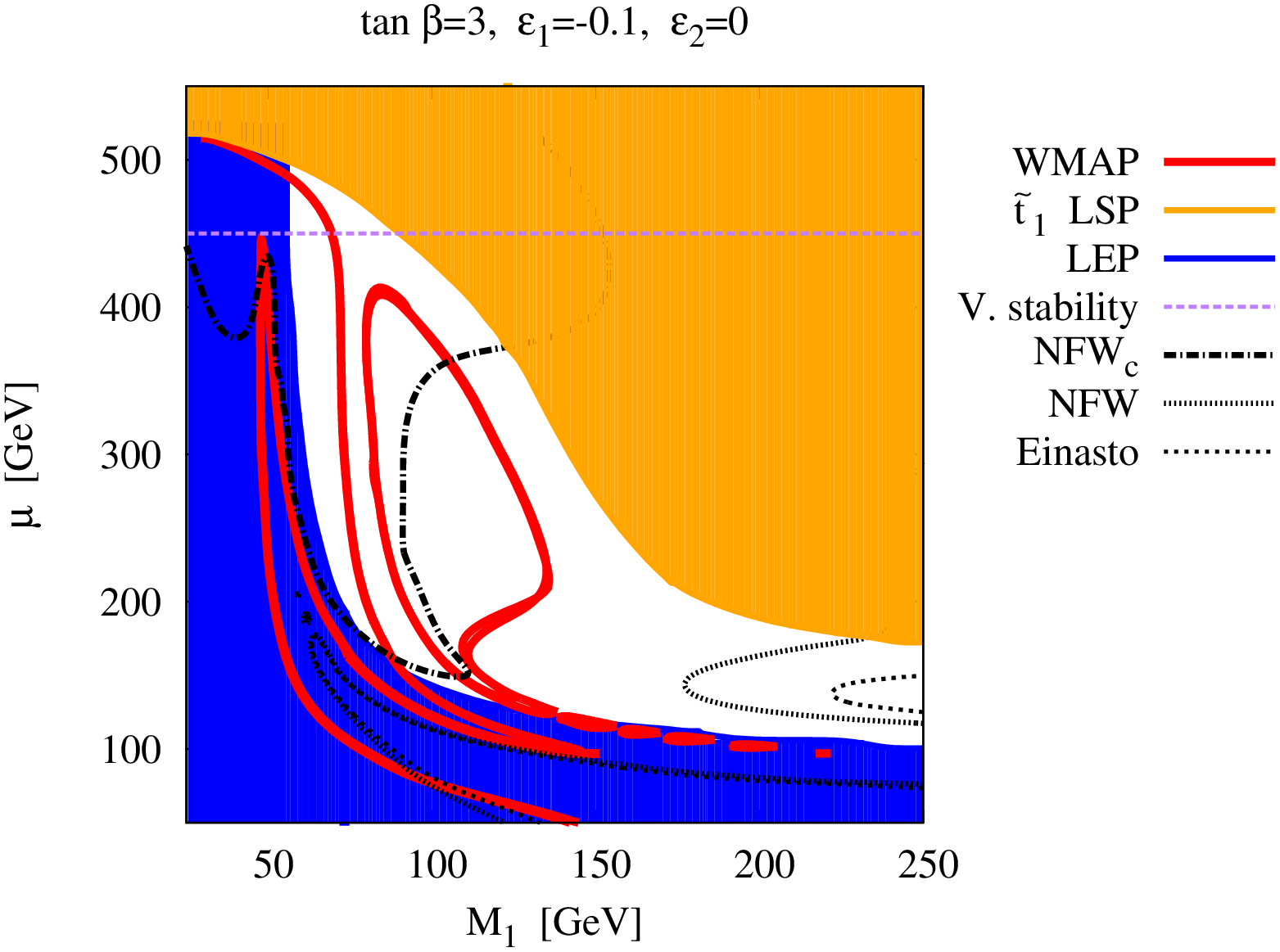,height=7cm,angle=-90}\vspace{-.3cm}
\caption{Regions in the $[m0,\,m_{1/2}]$ plane (upper panels) or $[M_1,\,\mu]$ plane (lower panels) in which the WMAP constraint is fulfilled (red lines).
The blue region is excluded by the null searches for charginos at LEP.
The oranges area in the upper (lower) panels are excluded because the stau (stop) is the LSP.
The dash-dotted black curves in the up panel are contour lines for the Higgs mass with values in GeV as indicated.
Above the dotted purple line of the left of this line, the electroweak vacuum is metastable.
\label{indirect}}
\end{center}
\end{figure}

In figure \ref{indirect} we present the detectability regions for the two models described previously, in the $[m_0,\,m_{1/2}]$ (upper panels) and $[M_1,\,\mu]$ (lower panels) parameter space.
FERMI will be sensitive to the regions below the contours and to the areas inside the blobs.
The detection prospects are maximised for low values of the pairs $m_0$
and $m_{1/2}$ or $M_1$ and $\mu$. For the mSUGRA-like model (upper panels), the growth of $m_{1/2}$
gives rise to the opening of some relevant production channels,
after passing some thresholds, increasing significantly $\langle\sigma v\rangle$.
The first one corresponds to a light neutralino, with mass $m_\chi\sim m_Z/2$
($m_{1/2}\sim 130$ GeV). In that case the annihilation is done via the $s$-channel exchange
of a real $Z$ boson.
The second threshold appears for $m_\chi\sim m_W$ ($m_{1/2}\sim 220$ GeV). The annihilation cross-section
is enhanced by the opening of the production channel of two real $W^\pm$ bosons in the final state.
The last threshold corresponds to the opening of the channel $\chi_1^0\chi_1^0\to t\bar{t}$
($m_{1/2}\sim 400$ GeV).
The aforementioned threshold appears as a particular feature on the upper plots:
an isolated detectable region for $m_{1/2}\sim 400$-$500$ GeV and $m_0\lesssim 300$ GeV.
Furthermore, larger values for the annihilation cross-section can be reached for higher values of $\tan\beta$.
For the second benchmark (lower panels), the region where $M_1\gg\mu$ is highly favored for indirect detection due to the fact that the LSP is higgsino-like,
maximising its coupling to the $Z$ boson. Let us recall that the $Z$ boson does not couple to a
pure gaugino-like neutralino.
The introduction of the NR operators gives rise to a mild signature when the neutralino is almost a bino-like state.
However, when $\mu <M_1$ there is an important increase of the $\chi^0_1-\chi^0_1-A$ coupling, and therefore to a boost in the annihilation into fermion pairs. On the other hand, as the Higgs boson $h$ becomes heavier, the processes giving rise to the final state $h\,Z$ get kinematically closed.
Concerning figure \ref{indirect}, let us note that the
only astrophysical setup in which some useful information can be extracted
is the NFW$_c$ one. This means that in this scenario, in order to have some positive detection
in the $\gamma$-ray channel, there should exist some important enhancement of the signal by some
astrophysical mechanism (as the adiabatic contraction mechanism invoked in this
case).

\section*{Acknowledgments}
I would like to thank the organising committee for inviting me at this pleasant conference.
I also want to thank the valuable collaboration with K. Blum, A. Goudelis, M. Losada and Y. Nir.
This work is supported by the E.C. Research Training Networks under contract MRTN-CT-2006-035505 HEPTools.


\begin{thebibliography}{}

\end{thebibliography}


\section*{References}
\begin{thebibliography}{99}
\bibitem{Djouadi:2005gj}
  A.~Djouadi,
  Phys.\ Rept.\  {\bf 459} (2008) 1
  [arXiv:hep-ph/0503173].
\bibitem{BMSSM}
  A.~Brignole, J.~A.~Casas, J.~R.~Espinosa and I.~Navarro,
  Nucl.\ Phys.\  B {\bf 666} (2003) 105
  [arXiv:hep-ph/0301121].
  M.~Dine, N.~Seiberg and S.~Thomas,
  Phys.\ Rev.\  D {\bf 76} (2007) 095004
  [arXiv:0707.0005 [hep-ph]].
  I.~Antoniadis, E.~Dudas, D.~M.~Ghilencea and P.~Tziveloglou,
  Nucl.\ Phys.\  B {\bf 808} (2009) 155
  [arXiv:0806.3778 [hep-ph]].
\bibitem{Blum:2009na}
  K.~Blum, C.~Delaunay and Y.~Hochberg,
  Phys.\ Rev.\  D {\bf 80} (2009) 075004
  [arXiv:0905.1701 [hep-ph]].
\bibitem{Dunkley:2008ie}
  J.~Dunkley {\it et al.}  [WMAP Collaboration],
  Astrophys.\ J.\ Suppl.\  {\bf 180} (2009) 306
  [arXiv:0803.0586 [astro-ph]].
\bibitem{Berg:2009mq}
  M.~Berg, J.~Edsj\"o, P.~Gondolo, E.~Lundstr\"om and S.~Sj\"ors,
  JCAP {\bf 0908} (2009) 035
  [arXiv:0906.0583 [hep-ph]].
\bibitem{Bernal:2009hd}
  N.~Bernal, K.~Blum, Y.~Nir and M.~Losada,
  JHEP {\bf 0908} (2009) 053
  [arXiv:0906.4696 [hep-ph]].
\bibitem{Bernal:2009jc}
  N.~Bernal and A.~Goudelis,
  JCAP {\bf 1003} (2010) 007
  [arXiv:0912.3905 [hep-ph]].
\bibitem{baryogenesis}
  K.~Blum and Y.~Nir,
  Phys.\ Rev.\  D {\bf 78} (2008) 035005
  [arXiv:0805.0097 [hep-ph]].
  K.~Blum, C.~Delaunay, M.~Losada, Y.~Nir and S.~Tulin,
  arXiv:1003.2447 [hep-ph].
\bibitem{micromegas}
  G.~B\'elanger, F.~Boudjema, A.~Pukhov and A.~Semenov,
  Comput.\ Phys.\ Commun.\  {\bf 176} (2007) 367
  [arXiv:hep-ph/0607059].
  G.~B\'elanger, F.~Boudjema, A.~Pukhov and A.~Semenov,
  Comput.\ Phys.\ Commun.\  {\bf 180} (2009) 747
  [arXiv:0803.2360 [hep-ph]].
\bibitem{Djouadi:2002ze}
  A.~Djouadi, J.~L.~Kneur and G.~Moultaka,
  Comput.\ Phys.\ Commun.\  {\bf 176} (2007) 426
  [arXiv:hep-ph/0211331].



\end{thebibliography}
\end{document}